\documentstyle{mn}

\newif\ifAMStwofonts

\ifoldfss
  \ifCUPmtlplainloaded \else
    \NewTextAlphabet{textbfit} {cmbxti10} {}
    \NewTextAlphabet{textbfss} {cmssbx10} {}
    \NewMathAlphabet{mathbfit} {cmbxti10} {} 
    \NewMathAlphabet{mathbfss} {cmssbx10} {} 
  \fi
  \ifAMStwofonts
    \ifCUPmtlplainloaded \else
      \NewSymbolFont{upmath} {eurm10}
      \NewSymbolFont{AMSa} {msam10}
      \NewMathSymbol{\upi}     {0}{upmath}{19}
      \NewMathSymbol{\umu}     {0}{upmath}{16}
      \NewMathSymbol{\upartial}{0}{upmath}{40}
      \NewMathSymbol{\leqslant}{3}{AMSa}{36}
      \NewMathSymbol{\geqslant}{3}{AMSa}{3E}

    \fi
  \fi
\fi 

\ifnfssone
  \newmathalphabet{\mathit}
  \addtoversion{normal}{\mathit}{cmr}{m}{it}
  \addtoversion{bold}{\mathit}{cmr}{bx}{it}
  \newmathalphabet{\mathbfit} 
  \addtoversion{normal}{\mathbfit}{cmr}{bx}{it}
  \addtoversion{bold}{\mathbfit}{cmr}{bx}{it}
  \newmathalphabet{\mathbfss} 
  \addtoversion{normal}{\mathbfss}{cmss}{bx}{n}
  \addtoversion{bold}{\mathbfss}{cmss}{bx}{n}
  \ifAMStwofonts
    \ifCUPmtlplainloaded \else
      \UseAMStwoboldmath
      \makeatletter
      \new@mathgroup\upmath@group
      \define@mathgroup\mv@normal\upmath@group{eur}{m}{n}
      \define@mathgroup\mv@bold\upmath@group{eur}{b}{n}
      \edef\UPM{\hexnumber\upmath@group}
      \new@mathgroup\amsa@group
      \define@mathgroup\mv@normal\amsa@group{msa}{m}{n}
      \define@mathgroup\mv@bold\amsa@group{msa}{m}{n}
      \edef\AMSa{\hexnumber\amsa@group}
      \makeatother
      \mathchardef\upi="0\UPM19
      \mathchardef\umu="0\UPM16
      \mathchardef\upartial="0\UPM40
      \mathchardef\leqslant="3\AMSa36
      \mathchardef\geqslant="3\AMSa3E
    \fi
  \fi
\fi 

\ifnfsstwo
  \DeclareMathAlphabet{\mathbfit}{OT1}{cmr}{bx}{it}
  \SetMathAlphabet\mathbfit{bold}{OT1}{cmr}{bx}{it}
  \DeclareMathAlphabet{\mathbfss}{OT1}{cmss}{bx}{n}
  \SetMathAlphabet\mathbfss{bold}{OT1}{cmss}{bx}{n}
  \ifAMStwofonts
    \ifCUPmtlplainloaded \else
      \DeclareSymbolFont{UPM}{U}{eur}{m}{n}
      \SetSymbolFont{UPM}{bold}{U}{eur}{b}{n}
      \DeclareSymbolFont{AMSa}{U}{msa}{m}{n}
      \DeclareMathSymbol{\upi}{0}{UPM}{"19}
      \DeclareMathSymbol{\umu}{0}{UPM}{"16}
      \DeclareMathSymbol{\upartial}{0}{UPM}{"40}
      \DeclareMathSymbol{\leqslant}{3}{AMSa}{"36}
      \DeclareMathSymbol{\geqslant}{3}{AMSa}{"3E}
    \fi
  \fi
\fi 

\ifCUPmtlplainloaded \else
  \ifAMStwofonts \else 
    \def\upi{\pi}
    \def\umu{\mu}
    \def\upartial{\partial}
  \fi
\fi

\title{The boundary layer of VW Hyi in quiescence} 
\author[P. Godon and E.M. Sion]
       {P. Godon and E.M. Sion  \\
       Villanova University, Villanova, PA 19085, USA} 
\date{Accepted . 
      Received ;
      in original form }

\pagerange{\pageref{firstpage}--\pageref{lastpage}}
\pubyear{2004}

\begin{document}

\maketitle

\label{firstpage}

\begin{abstract}

In this letter, we suggest that the missing boundary layer luminosity
of dwarf novae in quiescence is released mainly in the ultraviolet
(UV) as the second component commonly identified in the far ultraviolet
(FUV) as the "accretion belt".

We present the well-studied SU UMa-type system VW Hyi in detail
as a prototype for such a scenario.
We consider detailed multiwavelength observations
and in particular the recent FUSE observations of VW Hyi
which confirm the presence of a second component (the "accretion belt")
in the FUV spectrum of VW Hyi in quiescence.
The temperature ($\approx 50,000$K) and rotational velocity
($> 3,000$km~s$^{-1}$) of this second FUV component
are entirely consistent with the optically thick region
($\tau \approx 1$) located
just at the outer edge of optically thin boundary layer
in the simulations of Popham (1999).

This second component contributes 20\% of the FUV flux,
therefore implying a boundary layer luminosity:
$L_{BL} = 2 \times (0.2 \times L_{UV} + L_{X-ray}) =
0.6 \times L_{disc}$, while the theory (Klu\'zniak 1987) 
predicts, for the rotation rate of VW Hyi's WD,
$L_{BL} \approx 0.77 L_{disc}$.
The remaining accretion energy ($<0.1 L_{acc}$) is
apparently advected into the star as expected for optically
thin advection dominated boundary layers.
This scenario is consistent with the recent
simultaneous X-ray and UV observations of VW Hyi by 
(Pandel, C\'ordova \& Howell 2003), 
from which we deduced here that the alpha viscosity parameter
in the boundary layer region must be as small as
$\alpha \approx 0.004$.

\end{abstract}


\section{Introduction: VW Hyi
- a key system for understanding Dwarf Novae}

Dwarf novae (DNe) are mass-exchanging binaries
in which a low-mass main sequence-like star (the secondary)
fills its Roche lobe and loses hydrogen-rich matter
to a white dwarf (WD) star (the primary).
The mass transfer is regulated by an
accretion disc, which undergoes
cyclic changes of the mass accretion rate $\dot{M}$.
The low mass accretion rate
($\dot{M} \approx 10^{-11} M_{\odot}$~yr$^{-1}$)
quiescent stage is interrupted intermittently every few weeks (to months)
by a high mass accretion rate
($\dot{M} \approx 10^{-8} M_{\odot}$~yr$^{-1}$),
the DN outburst stage which lasts days (to weeks).

At 65 pc VW Hyi is one of the closests (Waner 1987) 
and brightest
example of the SU UMa sub-class of DNe,
which undergo
both normal DN outbursts and superoutbursts.
The mass of the accreting
WD was estimated to be 0.63 $M_{\odot}$ (Schoembs \& Vogt 1981),
but more recently a gravitational redshift determination
yielded a larger mass $M_{wd}=0.86 M_{\odot}$ 
(Sion et al. 1997).
The inclination of the system is $i \approx 60$ degrees
(Schoembs \& Vogt 1981)
and, with an orbital period of 107 minutes,
it lies below the CV period gap,
where gravitational wave emission is
thought to drive mass transfer, resulting in very low accretion
rates during dwarf nova quiescence.
In addition,
it lies along a line of sight with an ideally low H{\small{I}}
interstellar column of $\approx 6 \times 10^{17}$ cm$^{-2}$
(Polidan, Mauche \& Wade 1990). 
As a consequence,
VW Hyi has been observed in nearly all wavelength ranges,
and it is, therefore, one of the best-studied systems.

However, there have been some discrepancies between the
observed X-ray luminosity and the expected boundary layer (BL)
luminosity in VW Hyi and other systems.
We suggest here that the missing boundary layer luminosity
in quiescence is released mainly in the ultraviolet
as the second FUV component commonly identified
as the "accretion belt".

In the next section we review multi-wavelength observations
of VW Hyi and identify each components in the system.
In particular, in section 3, we suggest that the second FUV component
is the optically thick region at the outer edge the optically thin
boundary layer. We discuss the implications of this suggestion for
VW Hyi in detail in section 4, and we conclude this letter in section 5.

\section{Multiwavelength Observations of VW Hyi}

\subsection{The Disc in the Optical}

In quiescence VW Hyi has an optical magnitude of about 13.8,
and an optical flux of about
$F_{opt}=8.5 \times 10^{-11}$erg~cm$^{-2}$s$^{-1}$
(van Amerongen et al. 1987, Pringle et al. 1987). 
Because of its low mass accretion rate and temperature
($ < 8,000$K), the accretion disc is
expected to be the main source of of the optical flux
in quiescence, namely
$F_{disc}=F_{opt} = 8.5 \times 10^{-11}$erg~cm$^{-2}$s$^{-1}$.
The energy, dissipated and radiated locally in the disc,
is exactly  half of the accretion energy
(Shakura \& Sunyaev 1973, Lynden-Bell \& Pringle 1974): 
\[ L_{disc}= \frac{L_{acc}}{2}=\frac{G M_{wd} \dot{M}}{2 R_{wd}} ,  \]
where $G$ is the gravitational constant,
$M_{wd}$ is the mass of the WD, $R_{wd}$ is the radius of the WD
and  $\dot{M}$ is the mass accretion rate.

\subsection{The Boundary Layer in the X-ray}

The remaining available
accretion energy, in the form of rotational kinetic
energy, is expected to be dissipated
in the so called {\it{boundary layer}} (BL) - the interface between the
inner edge of the fast rotating Keplerian disc and the slowly
rotating surface of the accreting WD.
This energy amounts (Klu\'zniak 1987):
\[
\frac{L_{BL}}{L_{disc}} =
\left( 1 - \frac{V_{wd}}{V_K(R_{wd})} \right) ^2,
\]
where
$V_{wd}$ is the
(equatorial) rotational velocity of the WD surface and
$V_K(R_{wd})$ is the Keplerian speed at one stellar radius.
For VW Hyi we have $V_{wd} \sin{i} = 400$km~s$^{-1}$
and we set $V_K(R_{wd}) \sin{i} \approx 3,200$km~s$^{-1}$
(Godon et al. 2004), 
this leads to a ratio $L_{BL}/L_{disc}=0.77$.
For a non-rotating WD one has
$L_{BL}=L_{disc}=\frac{1}{2}L_{acc}$.
Because of its small radial extent,
the BL is expected to be very hot
(with a temperature $T_{BL}\approx 10^{8}$K) and
optically thin during quiescence, as the density there is very low.
This tiny component is therefore expected to emit basically
the other half of the accretion energy in the X-ray band.

X-ray observations of VW Hyi in quiescence first carried out
with EXOSAT and ROSAT
(van der Woerd \& Heise 1987, Belloni et al. 1991 - 
using a single temperature plasma)
revealed an X-ray bolometric flux
$F_{X-ray} \approx 1.5-1.9 \times 10^{-11}$erg~cm$^{-2}$s$^{-1}$.
Subsequent X-ray observations
with ASCA and XMM-Newton (using two- and multiple-temperature
plasma models 
(Hasenkopf \& Eracleous 2002, Pandel et al. 2003 - respectively)
revealed a total X-ray bolometric flux
smaller by a factor of about 2:
$F_{X-ray} \approx 5-8 \times 10^{-12}$erg~cm$^{-2}$s$^{-1}$.
The X-ray observations all revealed
a temperature $kT \sim$ a few keV, and possibly as high as
$kT \approx 6-8$keV.
Pandel et al. (2003) 
fit the line profile assuming that the X-ray emitting region
is a thin equatorial belt near the surface of the WD with a rotational
velocity $v \sin{i} = 540$ km~s$^{-1}$.

However, so far the X-ray observations for VW Hyi
have revealed a much smaller BL luminosity than expected:
$L_{BL} \approx 0.1 L_{disc}$ (Belloni et al. 1991),
while from geometrical consideration, namely assuming that the star
occults half of the BL, Pandel et al. (2003) 
obtained $L_{BL}/L_{disc} = 0.2$.

\subsection{The WD in the Ultraviolet}

In general the WD in DNe has a typical temperature of about
$T_{wd} \approx 15-50,000$K, and
it is expected to dominate the ultraviolet (UV) light in most
DNe in quiescence: $L_{UV}=L_{wd}$.
For VW Hyi in quiescence,
early IUE Observations (Mateo \& Szkody 1984) 
revealed that the
UV light from the system was dominated by the WD with
$T_{eff} = 18,000 \pm 2,000$K .
Verbunt et al. (1987) and Pringle et al. (1987)
later estimated that the flux
observed at IUE wavelengths was about the same as the one
observed at optical wavelengths, namely $F_{UV}=F_{opt}=
8.5 \times 10^{-11}$erg~cm$^{-2}$s$^{-1}$.
Much higher S/N spectra were later obtained
with {\it{HST/STIS}} by Sion et al. (1995, 1996, 2001), 
who confirmed the basic shape of the spectrum.
They found that the WD had a temperature of about 20,000K
(which varied by at least 2,000 K, depending on the
time since outburst)
with a rotation rate of about $\sim$ 400 km~s$^{-1}$.

\subsection{The Second Component in the Far Ultraviolet}

However, there have been some indications of an additional component
besides the white dwarf in the FUV spectrum of VW Hyi
and other DNe in quiescence
(e.g. the presence of emission lines and the bottoms of Lyman
alpha profiles which do not go to zero as in pure white dwarf).
While the dominant component is that of a WD,
the second component is a rather flat continuum with an effective
temperature that is much higher than that of the WD.
Long et al. (1993) 
first suggested a fast rotating hot accretion belt
around the WD as a second component
to fit {\it{HUT}} observations of U Gem.
Long et al. (1993) remarked
that the physical basis for an accretion belt might be
the spin-up of the surface layers of the WD during outburst
and the slow conversion of kinetic energy to heat as a result
of viscous heating in the differentially rotating atmosphere
(Kippehahn \& Thomas 1978, Kutter \& Sparks 1987, 1989).  

The presence of the accretion belt was confirmed later
for VW Hyi from its {\it{HST/STIS}} spectra 
(Sion et al. 1995, 1996 \& 2001) 
and from its {\it{IUE}} spectra 
(G\"ansicke \& Beuermann 1996). 
It was found
that the second component contributed about 20\% of the FUV flux
(with the WD contributing the remaining 80\%)
and remains pretty much the same 5 days apart
(Sion et al. 2001). 

More recently, a Far Ultraviolet Spectroscopic Explorer
({\it{FUSE}}) spectrum of VW Hyi was taken during
quiescence 
(Godon et al. 2004), 11 days after outburst.
With a usable wavelength range of 904-1188 \AA ,
{\it{FUSE}} is able to probe the wavelength range
$\lambda < 1150$ \AA (while both {\it{STIS \&  IUE}} have
$\lambda > 1150$ \AA ) where the 20,000K WD
is not expected to contribute to the spectrum,
therefore making it possible to unambiguously decide on the
nature of the second component.
The best-fitting model to the {\it{FUSE}} data was a composite
model consisting of a 23,000K WD
rotating at $V_{rot}\sin{i}= 400$km~s$^{-1}$ with
a $\approx 48,000-50,000$K accretion belt rotating at a
velocity of $V_{belt}\sin{i} > 3,000$km~s$^{-1}$.
In this model, the white dwarf contributed 83\% of
the FUV flux and the accretion belt 17\% of the FUV flux.

\section{The Nature of the Second FUV Component}

The picture of the second component
that emerges from all the FUV observations
of VW Hyi is that of a fast rotating hot layer
contributing about 20\% of the total FUV flux.
However, if this component is an accretion belt formed
during outburst, then one expects
its velocity and temperature to decrease
during quiescence. However, observations show
that this accretion belt remains pretty much the same
during quiescence, whether it is observed a few days
or two weeks after outburst. This has led us to look for
a different interpretation of the fast rotating hot
component.

The only other region in the system that has
such a high velocity and temperature is just at the interface
between the inner edge of the disc and the outer edge of the
boundary layer, as shown in the simulations of the optically
thin BL during quiescence 
(Narayan \& Popham 1993, Popham 1999). 
These simulations show that in the boundary layer itself
the central temperature is very high
($\approx 10^8$K) but the effective temperature is low because
of the low optical depth.
However, except for the X-ray emitting
region directly adjacent to the stellar
surface, the only other region in the BL where the effective
temperature is high is in a small region just
outside the boundary layer,
where the optical depth is of the order of one.
This region coincides with a density increase
at the inner edge of the disc at the so-called transition
radius $R_{tr}$.
For $r<R_{tr}$ the BL is optically thin,
for $r>R_{tr}$ the disc is optically thick, and around
$r\approx R_{tr}$ one has an optical depth $\tau \approx 1 $.
For an accretion rate
$\dot{M} = 3.16 \times 10^{-11}M_{\odot}$yr$^{-1}$,
Popham (1999) found $R_{tr}=1.4 R_{wd}$ with a maximum temperature
in the transition region reaching about 60,000K.
The temperature and velocity of
this transition region are both in agreement with
the observations of the second
component in the FUV spectrum of VW Hyi in quiescence,
therefore implying that the transition region is probably the
source emitting the second FUV component. 
In the one-dimensional simulations of Popham (1999),
this second component forms a hot ring at the inner 
edge of the disc.  

Further evidence to support such a scenario was advanced
recently by Pandel et al. (2003)  who carried out
simultaneous X-ray and UV observations of VW Hyi.
They found that the variability of the X-ray and UV bands
are correlated
and that the X-ray flux is delayed by about 100 s over the
UV flux. They too suspected that the UV flux
is emitted near the BL (at $r\approx R_{tr}$)
and pointed out that accretion rate fluctuations
in this UV region are propagated in the X-ray emitting part of the BL
within 100 s. However, no temperature
or velocity of the emitting UV source was derived in Pandel et al. (2003)
and therefore the values of  $T\approx 50,000$K and
$V \sin{i} > 3,000$km~s$^{-1}$  derived in 
Godon et al. (2004) for the
second FUV component provides solid evidence to support that scenario
in which the boundary layer emits a substantial fraction
of its energy in the FUV.

It is important here to remark, however, that the
one-dimensional 'classic picture', in which the boundary layer
is treated as part of the disc, does actually break down
when the boundary layer becomes optically
thin and advection dominated. The geometry of the advection dominated
accretion flow changes dramatically from being thin and flattened
to being approximately spherical 
(Popham 1999, Popham \& Sunyaev 2001 - 
where the disc thickness $H$ becomes large: $H/r \approx 1$). 
In that case the boundary layer has to be treated as part
of the star rather than part of the disc, as pointed out
by Inogamov \& Sunyaev (1999), who carried out such a treatment
for accretion on to a neutron star.
In this approach,
the accreting gas reaches the equatorial region of the star
spinning at the Keplerian velocity and forms a layer on the
stellar surface which spreads from the equator toward the poles
and form a so-called  "spread layer" rather than a "boundary layer".
The deceleration of Keplerian rotation and energy release
take place on the stellar surface in a latitude belt.
As the matter loses angular momentum, the centrifugal
force decreases, allowing the matter to move from the
equator toward the poles.
The combined effect of centrifugal force and radiation pressure
(dominant in the case of accretion on to a neutron star) gives
rise to two latitude rings of enhanced brightness which are symmetric
about the equator. Such a treatment has not been carried out for
accretion on to a WD in the optically thin regime 
and it is not clear whether two such rings would
form in that case. However, what is clear is that the optically
thin boundary layer has to be spherical and envelops the equatorial
region of the star (spreading to some extent toward the poles).
The X-ray are expected to be emitted closer to the stellar surface
while the UV is emitted just outside the BL, where it meets the
disc. The X-ray emitting region and the UV emitting region are separated
by the optically thin spherical BL of thickness
$\Delta R \approx 0.4 R_{wd}$ (Popham 1999). 

\section{Discussion: The Case of VW Hyi}

Here, we discuss the results obtained so far for
VW Hyi and assess the luminosity and the viscosity
parameter in the BL of VW Hyi.

\subsection{The Luminosity}
We suggest that the second component observed in the FUV spectrum
of VW Hyi, commonly identified as the accretion belt, is really
the outer edge of the {\it{spread}} boundary layer and
therefore its contribution in the FUV flux has to be taken into
account when computing the boundary layer luminosity. This belt
accounts for 20\% of the FUV flux, and consequently the boundary layer
luminosity that is observed is actually:
$$
L_{BL} = L_X + 0.2 \times L_{opt} = 0.3 \times L_{opt}, $$
where we have substituted $L_{X-ray}=0.1 L_{opt}$ 
(Belloni et al. 1991) and
$L_{UV}=L_{opt}$ (Pringle et al. 1987). 
And since one-half of the boundary layer is
occulted by the star, the total emission
from the boundary layer is twice the amount observed, such that one has:
$$ \frac{L_{BL}}{L_{disc}} = 0.6 ,$$
since it is assumed that $L_{disc}=L_{opt}$.
To summarize, from an expected
$0.77L_{disc}$, the boundary layer radiates $0.6 L_{disc}$,
one third of which is emitted in the X-ray and two third emitted
in the FUV (as predicted by Pandel et al. 2003).
The remaining energy is most probably advected
into the outer layer of the star (implying $L_{adv}=0.17L_{disc}$),
as expected for
advection dominated optically thin boundary layers solutions
(Narayan \& Popham 1993, Popham 1999). 
However, no quantitative results have been
presented for the optically thin branch of solutions of the boundary layer.
The only quantitative results for advection dominated boundary layers
are for the optically thick branch 
(Popham 1997, Godon 1997) which
predict a ratio $L_{adv}/L_{acc} \approx 0.1-0.2$ in rough agreement with
what we find here $L_{adv}/L_{acc} \approx 0.085 \approx 0.1$.

\subsection{The Viscosity}

X-ray observations of VW Hyi in outburst 
(van der Woerd \& Heise 1987)  
have revealed the presence of a 14s DNO, which has been
associated with the rotation period of the Keplerian flow in the
very inner disc. On the other hand, Pandel et al. (2003) found
that the time it takes for the matter to transit through the boundary
layer is about 100s, which is equivalent to the
time  it takes to spin down the matter from Keplerian speed to
a stellar rotational velocity.  We use here these two
basic time scales to assess the viscosity parameter in the
boundary layer of VW Hyi.

From simple hydrodynamical considerations, the
spindown (or spinup) time $\tau_{spin}$ of a rotating flow
is the geometric mean of the rotation time $\tau_{rot}$
and the viscous diffusion time
$\tau_{\nu}$ (Gill 1982): 
\begin{equation}
\tau_{spin} = \sqrt{ \tau_{rot} \times \tau_{\nu} },
\end{equation}
or equivalently:
\begin{equation}
\tau_{\nu} = \frac{ \tau_{spin}^2 }{ \tau_{rot} } .
\end{equation}
Using the value $\tau_{rot} \approx 14$s 
(van der Woerd \& Heise 1987) and
$\tau_{spin} \approx 100$s 
(Pandel et al. 2003), leads to $\tau_{\nu} \approx 667$s.

An estimate of the viscosity coefficient $\nu$ in the boundary layer
region can then be assessed using a basic scaling argument.
The straight viscous diffusion in a flow would spread the boundary effects
in a time $t$ outward to a distance $\delta$ given by
$\delta = \sqrt{\nu t}$ (Gill 1982).   
Therefore the viscous diffusion time $\tau_{\nu}$
is given by the simple
relation\footnote{This relation is also known in the
context of Numerical Methods, in Computational Astrophysics,
where it is used in viscous fluid dynamics for the numerical
integration to proceed faster than the viscous physical processes
in the flow. In that case the time step $\Delta t$ must satisfies
$\Delta t < \tau_{\nu}$, and where $\Delta R$ is the grid spacing;
usually this takes over the Courant
Friedrich condition for the numerical integration speed.}:
$$
\tau_{\nu} = \frac{ \Delta R^2}{\nu } ,
$$
where $\Delta R$ is the size of the boundary layer.
Inverting the relation, one obtains:
\begin{equation}
\nu \approx \frac{\Delta R^2}{\tau_{\nu}}
\approx 1.5 \times 10^{14}cm^2s^{-1},
\end{equation}
where we have assumed
$\Delta R \approx H \approx 0.4 R_{wd}$ 
(Popham 1999, Pandel et al. 2003) 
and $R_{wd} \approx 8 \times 10^{8}$cm. This value is in agreement
within one order of magnitude with the value expected
for the disc viscosity. For a thiner boundary
layer region of $\Delta R \approx 0.1 R_{wd}$, the viscosity
parameter is smaller by an order of magnitude
$\nu \approx 10^{13}$cm$^2$s$^{-1}$.
For a temperature of $10^8$K in the BL, the sound speed
is close to $c_s \approx 10^8$cm~s$^{-1}$ and one has
$$
\alpha = \frac{\nu }{ c_s H} \approx
\frac{ 1.5 \times 10^{14} } { 10^8 \times 0.4 \times 8 \times 10^8 }
\approx 0.004. $$
We find that in the BL region the viscosity parameter $\alpha$
is rather small.

For comparison, in the inner disc, at
$r\approx 2R_{wd} \approx 1.6 \times 10^9$cm,
where $T < 10,000$K, one has $H/r \approx 0.01$,
$c_s \approx 10^6$cm~s$^{-1}$ and $\nu = \alpha  c_s H \approx
\alpha \times 10^{13}$cm$^2$s$^{-1}$.

\section{Conclusion}
In this letter, we claim that the BL in VW Hyi radiates 2/3 of
its emission in the UV (as the component previously identified as the
accretion belt) and 1/3 in the X-ray, summing up to 0.6
of the disc luminosity. The remaining BL energy ($0.17 L_{disc}$)
is apparently advected into the star. We suggest that the BL
in most quiescent DN is probably radiating a large fraction of
the emission in the UV.

We also estimated that
the alpha viscosity parameter in the BL region of VW Hyi is as small
as $\alpha \approx 0.004$.
The present findings are important, since the
origin and nature of
the viscosity in the boundary layer are not known and might
be very different from that in the Keplerian disc.
The Balbus-Hawley instability 
(Balbus \& Hawley 1991)
is not expected
to work in the boundary layer since there the relation
$d \Omega ^2 / dR < 0$ does not hold, and
in addition the boundary layer is stable to the centrifugal instability
according to the Rayleigh  criterion $d(\Omega R^2)/dR > 0$ 
(Popham \& Sunyaev 2001).  
It has been remarked 
(Brandenburg et al. 1995, Popham \& Sunyaev 2001) 
that non linear
instabilities in Poiseuille and Couette flows,
previously investigated as the origin of the turbulence
in the disc itself 
(Zahn 1990, Dubrulle 1991),  
could apply in the boundary layer region, because there the flow is
forced by the boundary in a similar way.
(i.e. the flow between two co-rotating cylinders when the
outer cylinder rotates much faster than the inner one).
It has also been
argued that $\alpha$ should depend on the shear 
(Godon 1996, Abramowicz, Brandenburg \& Lasota 1996) 
which is much larger in the boundary layer than in the disc.

\section*{Acknowledgments}

\bsp
\label{lastpage}
\end{document}